\documentclass[pre,twocolumn]{revtex4-1}
\usepackage{graphicx}

\begin{document}

\title{Relating urban scaling, fundamental allometry, and density scaling}

\author{Diego Rybski}
\affiliation{Potsdam Institute for Climate Impact Research -- PIK, Member of Leibniz Association, P.O. Box 601203,
14412 Potsdam, Germany}
\email{ca-dr@rybski.de}

\date{\today{} -- \jobname}

\begin{abstract}
We study the connection between urban scaling, fundamental allometry (between city population and city area), and per capita vs.\ population density scaling. 
From simple analytical derivations we obtain the relation between the 3 involved exponents.
We discuss particular cases and ranges of the exponents which we illustrate in a ``phase diagram''.
As we show, the results are consistent with previous work.
\end{abstract}

\maketitle

\section{Introduction}

Urban scaling -- i.e.\ the non-linear power-law relation between an urban indicator and city population -- is considered as cornerstone of contemporary city science \cite{BettencourtW2010}.
However, by investigating solely the relation with population, implicitly the assumption of constant population density is made.

Here we discuss analytically the situation of non-constant density, specifically as resulting from a power-law between population and area. 
The problem has been addressed in \cite{RybskiRWFSK2016} but not discussed to a full extent.
Thus, we answer the question, how (i) urban scaling \cite{BettencourtLHKW2007} relates via the (ii) fundamental allometry between city population and city area \cite{BattyF2011} to (iii) per capita vs.\ population density scaling \cite{Gudipudi2015,NewmanK1989}.

We derive the relation between the 3 involved exponents and discuss various particular cases.
By means of a ``phase diagram'' we illustrate under which conditions the per capita vs.\ population density exponent is positive or negative, which -- depending on the considered quantity -- determines whether high- or low-density cities are more efficient.

\section{Derivations}

\emph{Urban scaling} relates an urban indicator and city size in terms of population \cite{BettencourtLHKW2007}
\begin{equation}
e \sim s^\gamma
\, ,
\label{eq:urban}
\end{equation}
where $e$ is the considered urban indicator in absolute terms, $s$ is the total city population, and $\gamma>0$ is the scaling exponent. 
Reported values of $\gamma$ are mostly within the range between $0.75$ and $1.25$.

The \emph{fundamental allometry} relates city population and city area, see \cite{BattyF2011} and references therein,
\begin{equation}
s \sim a^\delta
\, ,
\label{eq:fundallo}
\end{equation}
where $a$ is the city area and $\delta>0$ is another exponent, sometimes called allometric coefficient \cite{BattyF2011}.
Typically, $\delta$ is found in the range between $0.6$ and $1.4$.

Rewriting Eq.~(\ref{eq:fundallo}) we obtain
\begin{eqnarray}
s^{1/\delta} & \sim & a \nonumber \\
s^{1/\delta-1} & \sim & a/s \nonumber \\
s^{1-1/\delta} & \sim & s/a \nonumber \\
s^{\frac{\delta-1}{\delta}} & \sim & s/a \label{eq:density1} \\
s & \sim & (s/a)^{\frac{\delta}{\delta-1}} \label{eq:density} 
\, .
\end{eqnarray}
Moreover, transforming Eq.~(\ref{eq:urban}) leads to
\begin{equation}
e/s \sim s^{\gamma-1} \nonumber
\end{equation}
and with Eq.~(\ref{eq:density}) we obtain the \emph{per capita vs.\ population density scaling} relation
\begin{equation}
e/s \sim (s/a)^{\frac{\delta}{\delta-1}(\gamma-1)}
\, ,
\label{eq:res1}
\end{equation}
so that the per capita indicator, $e/s$, scales with density, $s/a$, to the exponent
\begin{equation}
\alpha=\frac{\delta}{\delta-1}(\gamma-1)
\, .
\label{eq:arel}
\end{equation}
E.g.\ \cite{Gudipudi2015} report $\alpha\approx -0.8$ for total CO2 emissions.

\subsection{Cases}

\begin{figure}[h]
\includegraphics[width=\columnwidth]{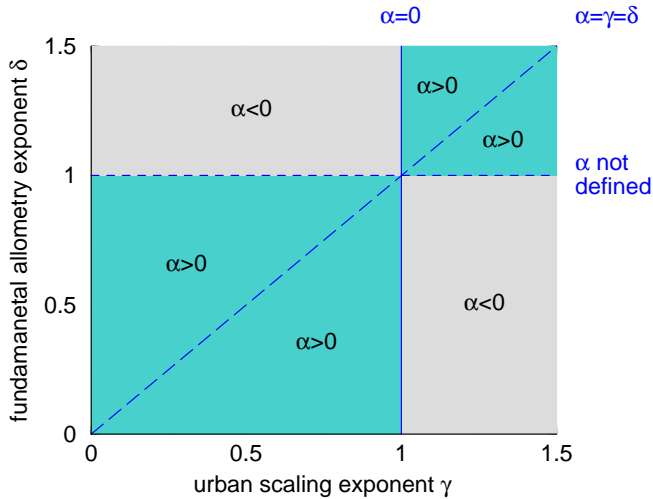}
\caption{\label{fig:space} ``Phase diagram'' illustrating particular cases and the sign of the per capita vs.\ population density exponent $\alpha$ from Eq.~(\ref{eq:essaa}) as obtained from Eq.~(\ref{eq:arel}) depending on the urban scaling exponent $\gamma$ from Eq.~(\ref{eq:urban}) and the fundamental allometry exponent $\delta$ from Eq.~(\ref{eq:fundallo}).
Straight lines indicate the particular cases and colored areas indicate whether $\alpha>0$ or $\alpha<0$.}
\end{figure}

From Eq.~(\ref{eq:arel}) the following cases can be distinguished.
\begin{itemize}
\item $\gamma=1$: $\alpha=0$, i.e.\ $e/s=\text{const.}$, the per capita indicator is independent of city density
\item $\delta=1$: $s/a=\text{const.}$, i.e.\ Eq.~(\ref{eq:essaa}) is not defined, this case is explicitely mentioned in \cite{OliveiraAM2014}
\item $\gamma=\delta=\alpha$: special case where all exponents are the same
\item $\gamma>1$, $\delta>1$: $\alpha>0$
\item $\gamma<1$, $\delta>1$: $\alpha<0$
\item $\gamma>1$, $\delta<1$: $\alpha<0$
\item $\gamma<1$, $\delta<1$: $\alpha>0$
\end{itemize}
The various parameter ranges and particular cases are illustrated in Fig.~\ref{fig:space}.

\subsection{Alternative derivation}

Alternatively, Eq.~(\ref{eq:arel}) can also be derived starting from the per capita vs.\ population density scaling relation
\begin{equation}
e/s \sim (s/a)^\alpha
\label{eq:essaa}
\end{equation}
and using Eq.~(\ref{eq:density1}) leads to
\begin{equation}
e/s \sim s^{\alpha\frac{\delta-1}{\delta}} \nonumber
\end{equation}
and finally to
\begin{equation}
e \sim s^{\alpha\frac{\delta-1}{\delta}+1} \nonumber
\, .
\end{equation}
Thus, comparison with Eq.~(\ref{eq:urban}) implies to $\gamma=\alpha\frac{\delta-1}{\delta}+1$ which is consistent with Eq.~(\ref{eq:arel}).

\section{Consistency with \cite[Rybski et al., Environ.\ Plan.\ B 2016]{RybskiRWFSK2016}}

Last we want to show that our results are consistent with the relations discussed in \cite{RybskiRWFSK2016}.
Rewriting Eq.~(\ref{eq:res1}) we obtain
\begin{equation}
e/s \sim (s/a)^\delta \, (s/a)^{\frac{\delta}{\delta-1}(\gamma-1)-\delta} \nonumber
\end{equation}
and simplifying the exponent
\begin{equation}
e/s \sim (s/a)^\delta \, (s/a)^{\delta\frac{\gamma-\delta}{\delta-1}} \nonumber
\, .
\end{equation}
Inserting Eq.~(\ref{eq:density1}) leads to
\begin{equation}
e/s \sim (s/a)^\delta \, s^{\gamma-\delta} \nonumber
\end{equation}
which corresponds to Eq.~(3) in \cite{RybskiRWFSK2016}.

\section{Discussion}

In summary, we have related 3 scaling laws which are commonly analyzed for cities.
The involved exponents are connected by a simple equation. 
The derivations are basic math but it might be useful to write them down once.

By means of a ``phase diagram'' we discuss various cases.
As one would expect, the per capita indicator is independent of the density if urban scaling resembles proportionality.
A special case occurs when all exponents are the same.
Whether the per capita indicator increases or decreases with density depends on whether the other two relations are super- or sub-linear.

Last, we would like to remark, that urban scaling in the form of Eq.~(\ref{eq:urban}) is criticized, see e.g.\ \cite{ArcauteHFYJB2014,LeitaoMGA2016} and references therein. 
Here, we do not have any position on such questions and simply assume the power-laws, Eqs.~(\ref{eq:urban}), (\ref{eq:fundallo}), and (\ref{eq:essaa}), hold true.

\vspace{-0.5cm}

\begin{acknowledgments}
We thank A.-L.\ Winz for valuable input.
The research leading to these results has received funding from the European
Community's Seventh Framework Programme under Grant Agreement No.~308497
(Project RAMSES).
\end{acknowledgments}

\bibliography{arXiv.bib}

\end{document}